# Ferromagnetic Resonance of a YIG film in the Low Frequency Regime

Seongjae Lee,[1] Scott Grudichak,[2] Joseph Sklenar,[2] C. C. Tsai,[3] Moongyu Jang,[4] Qinghui Yang,[5] Huaiwu Zhang,[5] and John B. Ketterson[2]

[1]Department of Physics, Research Institute for Natural Sciences, Hanyang University, Seoul, 133-791 South Korea

[2]Department of Physics and Astronomy, Northwestern University, Evanston IL, 60208 USA

[3]Department of Engineering & Management of Advanced Technology, Chang Jung Christian University, Tainan 71101, Taiwan

[4]Department of Materials Science and Engineering, Hallym University, Chuncheon, 200-702 South Korea

[5]State Key Laboratory of Electronic Films and Integrated Devices, University of Electronic Science and Technology, Chengdu, Sichuan, 610054, China

**Abstract**

An improved method for characterizing the magnetic anisotropy of films with cubic symmetry is described and is applied to an yttrium iron garnet (111) film. Analysis of the FMR spectra performed both in-plane and out-of-plane from 0.7 to 8 GHz yielded the magnetic anisotropy constants as well as the saturation magnetization. The field at which FMR is observed turns out to be quite sensitive to anisotropy constants (by more than a factor ten) in the low frequency (< 2 GHz) regime and when the orientation of the magnetic field is nearly normal to the sample plane; the restoring force on the magnetization arising from the magnetocrystalline anisotropy fields is then comparable to that from the external field, thereby allowing the anisotropy constants to be determined with greater accuracy. In this region, unusual dynamical behaviors are observed such as multiple resonances and a switching of FMR resonance with only a 1 degree change in field orientation at 0.7 GHz.

## Introduction

Yttrium iron garnet (YIG, $Y_3Fe_5O_{12}$) is a well-known ferromagnetic insulator with extremely low magnetic damping that has been widely used in microwave devices, and its spin wave properties have been extensively studied for decades.[1,2] Recently there have been important discoveries of spin-related phenomena such as spin pumping,[3,4,5] spin Seebeck effect,[6,7] and spin Hall magnetoresistance[8,9] in which YIG films played a central role. As an example, several authors have reported that the injection efficiency of a spin current from YIG to Pt is strongly enhanced at low resonant frequencies ($\leq$ 3 GHz) or low fields ($\leq$ 2 kG). [4,10-13] Thus it is important to have an accurate characterization of the magnetic response, in particular as it affects magnetization direction in YIG thin films; this is especially important at low frequencies and near-normal field orientations.

FMR is a powerful tool to probe the internal magnetic field of magnetic materials [14,15] and has become a standard technique to provide information about magnetocrystalline anisotropy, as well as the uniaxial anisotropy that typically emerges in thin films.[16] Several authors have reported high-quality YIG thin films prepared by pulsed laser deposition, [17,18,19] sputter deposition, [20] or chemical vapor deposition [21] along with FMR characterization. They employed a conventional FMR setup involving a narrow band cavity resonator where one sweeps the magnetic field through the resonance at different orientations but at a fixed frequency (typically ~9 GHz); the anisotropy constants are obtained by fitting the data to a theoretically predicted form. [18,19,21] The quantity $4\pi M$ is usually measured independently when obtaining the magnetic anisotropy constants. Since the required external field for the resonance is much higher than the anisotropy fields of YIG, the accuracy with which the anisotropy constants can be determined is limited.

In this paper, we present a new method to obtain the anisotropy constants of a YIG film with better accuracy by using a broadband FMR spectrometer in which resonant fields are measured not only as a function of the field orientation, as specified by both the polar angle, $\theta$, but also over frequencies that range from 0.7 GHz to 8 GHz. Our method has three advantages over those of angle-resolved FMR at fixed frequency[22,23]: 1) the magnetization value can be *directly extracted* from the frequency-dependent FMR data without resorting to independent magnetometry measurements; 2) the azimuthal angle of the field orientation with respect to the crystal axis is determined as a part of the fitting process, thereby eliminating any error associated with aligning sample; 3) the resonant field change resulting from a shift in the anisotropy constant turns out to be strongly amplified at low frequencies ($\leq 2$ GHz), by more than 10 times compared to that at 8 GHz. Furthermore, in this region we observe multiple resonances that can further enhance the accuracy. Finally, at a low frequency of 0.7 GHz it is observed that a 1 degree change in field orientation induces an abrupt switching phenomenon in the FMR resonance. All these behaviors follow from the Landau-Liftshitz theory as augmented by the inclusion of appropriate anisotropy energies. [24]

**Theoretical Background**

Figure 1 shows a schematic illustration of the coordinate system used for our (111) oriented epitaxial YIG film, where the film normal is taken as the z-axis and the polar angles and azimuthal angles for the effective magnetization $\mathbf{M}$ and an external magnetic field $\mathbf{H}$ are denoted as $(\psi, \phi_M)$ and $(\theta, \phi_H)$, respectively. The free energy density describing the magnetization of a ferromagnetic film having a cubic structure is given by [24]

$$F = -\mathbf{H} \cdot \mathbf{M} + 2\pi \mathbf{M} \cdot \mathbf{N} \cdot \mathbf{M} + K_1 \left( M_x^2 M_y^2 + M_y^2 M_z^2 + M_z^2 M_x^2 \right) + K_2 M_x^2 M_y^2 M_z^2 \quad (1)$$

Here the first term is the Zeeman energy, the second is the demagnetization energy involving a demagnetization tensor $\mathbf{N}$, the third and fourth terms represent the cubic first and second order magnetic anisotropy energies (MAE) characterized by constants $K_1$ and $K_2$ and $M_x$, $M_y$ and $M_z$ are the vector components of $\mathbf{M}$ along the crystal coordinates: x, y, z.

Assuming the film as an infinite plane sheet where $N_x = N_y = 0$, $N_z = 1$ and taking the z-axis and y-axis parallel to the [111] and [01$\bar{1}$] directions in the crystal coordinates, respectively, Eq. (1) becomes

$$F = -HM\{\cos\theta\cos\psi + \sin\theta\sin\psi\cos(\phi_H - \phi_M)\} + 2\pi M^2 \cos^2\psi - K_u \cos^2\psi$$

$$+ \frac{K_1}{12}\left(7\sin^4\psi - 8\sin^2\psi + 4 - 4\sqrt{2}\sin^3\psi\cos\psi\cos 3\phi_M\right)$$

$$+ \frac{K_2}{108}\Big(-24\sin^6\psi + 45\sin^4\psi - 24\sin^2\psi + 4$$

$$-2\sqrt{2}\sin^3\psi\cos\psi\left(5\sin^2\psi - 2\right)\cos 3\phi_M + \sin^6\psi\cos 6\phi_M\Big). \qquad (2)$$

Here we included the out-of-plane uniaxial anisotropy energy in the third term as characterized by a constant $K_u$. For conveniences in the analysis, we introduce dimensionless variables: $h \equiv \frac{H}{4\pi M}$, $k_u \equiv \frac{K_u}{2\pi M^2}$, $k_1 \equiv \frac{K_1}{2\pi M^2}$ and $k_2 \equiv \frac{K_2}{2\pi M^2}$ and then the above equation becomes

$$\frac{F}{2\pi M^2} = -2h\{\cos\theta\cos\psi + \sin\theta\sin\psi\cos(\phi_H - \phi_M)\} - (1-k_u)\cos^2\psi$$

$$+ \frac{k_1}{12}\left(7\sin^4\psi - 8\sin^2\psi + 4 - 4\sqrt{2}\sin^3\psi\cos\psi\cos 3\phi_M\right)$$

$$+ \frac{k_2}{108}\Big(-24\sin^6\psi + 45\sin^4\psi - 24\sin^2\psi + 4$$

$$-2\sqrt{2}\sin^3\psi\cos\psi\left(5\sin^2\psi - 2\right)\cos 3\phi_M + \sin^6\psi\cos 6\phi_M\Big) \qquad (3)$$

Note that the cubic anisotropy energy terms each result in a three-fold symmetry in the azimuthal angle $\phi_M$ for a (111) oriented film, leading to different resonant fields when $\theta \to -\theta$ as will be discussed below. We also note that the anisotropy terms contain higher order terms in $\psi$ and $\phi_M$, thus producing higher order contributions in their second derivatives when $\psi \sim 0$. Since the FMR frequency depends on the second derivatives, this term will play an important role in the angular dependence of the FMR response. For small $\theta$, the second derivative of Zeeman term with respect to $\phi_M$ is comparable with those of MAE term and cancels out, leading to a soft mode, even when the external field H is much larger than internal fields arising from MAE; i.e., $H_{A1}(\equiv 2K_1/M = 4\pi M k_1)$ and $H_{A2}(\equiv 2K_2/M = 4\pi M k_2)$. For bulk single crystal YIG, the anisotropy field values are $H_{A1} = -87.2$ Oe and $H_{A2} = -3.71$ Oe while $4\pi M = 1760$ gauss. [25]

Given the set of parameters h, $\theta$ and $\phi_H$ that define the external field, the equilibrium orientation of the magnetization, specified by $\psi$ and $\phi_M$, is determined by minimizing the free energy:

$$\frac{\partial F}{\partial \psi} = 0, \quad \frac{\partial F}{\partial \phi_M} = 0, \tag{4}$$

with the additional conditions of positive curvatures $\frac{\partial^2 F}{\partial \psi^2} > 0$ and $\frac{\partial^2 F}{\partial \phi_M^2} > 0$. A small orientational perturbation of the magnetization from this equilibrium position is counteracted by the restoring forces that depend on curvatures of the free energy, which dictate the dynamics of the magnetization. The general expression for the FMR frequency $f_{res}$ was derived by Suhl, Smit and Beljers in 1955: [26, 27]

$$f_{res} = \frac{\gamma}{2\pi M \sin\psi} \left[ \left(\frac{\partial^2 F}{\partial \psi^2}\right)\left(\frac{\partial^2 F}{\partial \phi_M^2}\right) - \left(\frac{\partial^2 F}{\partial \psi \partial \phi_M}\right)^2 \right]^{1/2} \tag{5}$$

where $\gamma$ is the gyromagnetic ratio.

There are no general analytical solutions of Eq. 5 for the FMR frequency. A special case occurs when the field is applied normal to the film; i.e., where $\theta = 0$. When $h \geq 1 - k_u + \frac{2k_1}{3} + \frac{2k_2}{9}$ the magnetization aligns with H ($\psi = 0$) and the FMR frequency behaves linearly with the field strength:

$$\frac{f_{res}}{2\gamma M} = h - \left(1 - k_u + \frac{2k_1}{3} + \frac{2k_2}{9}\right) \quad (\theta = 0) \tag{6}$$

Using Eq.'s (3) through (5) we numerically solved for the resonant frequency, $f_{res}(h, \theta)$ using $(k_u, k_1, k_2, \phi_H)$ as parameters that were then varied to obtain a least squares fit when compared with the experimental FMR data (this was accomplished using the commercially available MATLAB program); the quantity $4\pi M$ is obtained by comparing the data set for

$f_{res}(H, \theta = 0)$ with Eq. 6. Note we take the origin of the experimental azimuthal angle $\phi_H$ as a fitting parameter, so as to minimize its uncertainty relative to the [110] axis. [18]

**Experimental Methods**

The magnetic systems studied here are 6-μm thick yttrium iron garnet (YIG) single crystalline films of composition $Y_3Fe_5O_{12}$, grown by a liquid phase epitaxial technique on the (111) oriented $Gd_3Ga_5O_{12}$ (GGG) garnet substrate.[28, 29] Data were acquired with an FMR spectrometer capable of varying the frequency (f), DC magnetic field (h), and orientation (θ). It operates in a transmission mode with a sample cell placed in the rotatable Varian electro-magnet with maximum fields of ±6 kOe. The cell consists of a copper housing with input and output microwave coax connectors and an internal chamber that clamped our YIG film adjacent to a meander line formed from copper wire which generates the RF magnetic fields; details concerning the construction of the cell are presented elsewhere.[30] The microwave source is a HP model 83623A synthesizer with a frequency range of 10 MHz–20 GHz. The input RF signal was modulated in amplitude at a frequency of 4000 Hz and fed into the meanderline in contact with the YIG film. The output RF signal from the meanderline was rectified by a microwave diode and the resulting signal was sent to a lock-in amplifier while sweeping the DC magnetic field, which directly yields the FMR absorption line. For each θ, we carried out sweeps with increasing and decreasing magnetic field as well as positive and negative fields. The field was swept by a bipolar Kepco operational amplifier power supply driven from the analog output of a National Instruments D/A board inserted in a computer operating under software created in Labview; the same board digitized the lock-in output. When necessary, the signal to noise could be enhanced by collecting data in a signal averaging mode with a programmable number of sweeps.

**Results and Discussions**

The resonant modes of the YIG film were measured by sweeping the magnetic field at seven different frequencies 0.7, 1, 1.6, 2.5, 4, 6, and 8 GHz and fixed polar angles from −90 deg to +90 deg. The frequency vs. field behavior for the resonances observed with the field normal to the film ($\theta = 0$) is plotted in Fig. 2, and shows a linear behavior that is fitted as $f_{res} = 2.818\,(H - 1.612)$, as expected from theory (Eq. 6), where the units of $f_{res}$ and H are GHz and kOe. From the slope of this relation we obtain $\gamma/2\pi = 2.818\,\text{GHz/kOe}$, which is close to the free electron's value 2.803 GHz/kOe. From the intercept, we have an important relation between the saturation magnetization in kG and the terms of $k_u$, $k_1$, and $k_2$:

$$4\pi M = \frac{1.612}{1 - k_u + 2k_1/3 + 2k_2/9}. \tag{7}$$

Note the quantity $4\pi M$ of our sample is determined as soon as $k_u, k_1,$ and $k_2$ are fixed. As previously mentioned, it is important to obtain the magnetization value without resorting to independent measurements using a magnetometer, which can result in as much as a 5~10% error in extracting the anisotropy constants. [19]

Fig. 3 plots the resonant field (H) vs. angle ($\theta$) for seven frequencies with various symbols. At lower frequencies and with small $\theta$ we observed multiple resonances in field sweeps where we denote the highest resonant field as the primary resonance and refer to the others as secondary. For simplicity we display only the primary resonance field data in this figure. However, we took the data for all of the (multiple) resonance into account in our fit. The resulting parameters are: $k_u = -0.01143$, $k_1 = -0.05040$, $k_2 = 0.01276$, and $\phi_H = 4.599$. Theoretical curves using these values are displayed as the solid lines in the figure. The fit between theory and experiment is consistent for all frequencies. Substituting these three k-parameters into Eq. 7 yields $4\pi M = 1.644\,\text{kG}$, which is comparable to the bulk value 1.76 kG.[23] The corresponding conventional magnetic anisotropy constants (fields) are $K_1 = -5.421 \times 10^3\,\text{erg/cm}^3\,(H_{A1} = -82.87\,\text{Oe})$, $K_2 = 1.372 \times 10^3\,\text{erg/cm}^3\,(H_{A2} = 20.98\,\text{Oe})$, and $K_u = -1.230 \times 10^3\,\text{erg/cm}^3\,(H_u = -18.80\,\text{Oe})$. The first order anisotropy field $H_{A1} = -82.87\,\text{Oe}$ is in good agreement with previous work on bulk YIG, $-87.2\,\text{Oe}$ [25] or on thin films, which are centered on $-80\,\text{Oe}$. [17, 18, 19, 21] However the second order anisotropy field $H_{A2}\,(\equiv 2K_2/M)$ is about an order of magnitude higher than the bulk value of $-3.71\,\text{Oe}$ [25]. Judged from the two constants $K_1$ and $K_2$, we see that the easy axis for YIG film is in the (111) direction.

One important feature that follows from Fig. 3. is the asymmetry when $\theta \to -\theta$ due to the three-fold symmetry of the free energy in $\phi_M$. The asymmetric behavior of resonant fields is more pronounced as we lower frequency. In the inset, we show the data at some lower frequencies 0.7, 1, 1.6 GHz at near-normal angles to emphasize this asymmetric nature of the FMR spectra. This asymmetry in $\theta$ improves the accuracy in fitting the data. In particular at 1 GHz, a sudden drop of resonant field from 1.878 kOe to 0.562 kOe is observed as $\theta$ changes from $-2°$ to $-4°$ that is contrasted from the smooth behavior at the corresponding positive angles. Note that this behavior follows from the theory, as seen by the solid lines. We emphasize again that this abruptly changing behavior is only apparent in a low frequency regime, $\leq 1$ GHz. For the case of 0.7 GHz, data are absent except for several small angles near zero because resonance fields are so low that domain dynamics prevail over uniform magnetization precession.

In order to investigate the sensitivity of the data to the fitted anisotropy constants, we computed the fractional change in the resonant field when $k_1$ is shifted by 5% from the best-fit value, $k_1^{(0)} = -0.05040$,

$$\left[ H_{res}(k_1^{(0)} + 0.05 k_1^{(0)}) - H_{res}(k_1^{(0)}) \right] / H_{res}(k_1^{(0)}),$$

for various frequencies and angles; the results are shown in Fig. 4. At the highest frequency 8 GHz, the most sensitive orientation is $\theta \approx -45°$ with the sensitivity ~ 0.2 %, but as frequency is lowered, the polar angle of the most sensitive orientation increases gradually. For 1.6 GHz, the sensitivity rises up to 3 % at $\theta \approx -6.8°$, which is more than ten times larger than for 8 GHz, corresponding to an increase in resulting accuracy. This increase in the sensitivity at lower frequencies for near-normal orientations of external field is due to the fact that the second derivative of magnetocrystalline energy term with respect to $\phi_M$ is comparable to that of Zeeman energy. In addition, there is a sign change when passing through $\theta = 0$; this is related to the asymmetry mentioned above, which enhances the accuracy of the fitting. The sensitivity of the resonant field to a 5% deviation of $\phi_H$ is also computed in the same manner and is shown in the inset where the maximum of an ~ 2 % change occurs near $\theta \cong +5.5°$ at 0.7 GHz. This degree of sensitivity allows for a pinpoint determination of the azimuthal angle of the field; precise alignment of experimental apparatus with crystal axis is not required, thereby avoiding the error associated with it.

The asymmetric feature due to MAE stands out best in the frequency vs field plot of FMR data together with theoretical curves for fixed angles, as shown in Fig. 5 in which we restrict the data to the primary resonance and the angular range $-20° \leq \theta \leq 20°$ for clear viewing. Here the FMR data and theoretical curves for positive angles are denoted by filled symbols and solid lines respectively and those for negative angles by open symbols with dotted lines. We can see the soft mode behaviors in the low frequency range (<1.2 GHz) with near normal orientation ($|\theta| \leq 4°$). It is attributed to the fact that the restoring forces on the magnetization in azimuthal direction cancels out for small θ; that is, the second derivatives of Zeeman term with respect to $\phi_M$ is comparable to that of MAE term in the free energy. From this figure, we can clearly note the difference between branches with positive and negative polar angles. For a given angle, the positive branch bends downward more rapidly than the negative branch near $H \sim 4\pi M$. At 1 GHz, for example, the resonant field is 1.729 kOe for $\theta = +4°$ which is to be contrasted with 0.562 kOe for $\theta = -4°$. We can also easily understand the rapid drop in $H_{res}$ in the inset of Fig. 3 where we see the local minimum point is lifted up above 1 GHz as θ decreases from $-2°$ to $-4°$. It is noteworthy that for small θ, the soft mode feature has a local minimum which leads to *multiple resonant fields* for frequencies less than ~1.2 GHz. In this region, the FMR spectrum is very sensitive to either the frequency or the angle, an example being the curve for $\theta = -2°$ denoted by a dotted blue line in Fig. 5. At 1 GHz, we have multiple resonances. If we lower the frequency to 0.7 GHz, the resonance at high field disappears and only one resonance, at a low field 0.285 kOe, survives since the minimum point near 1.7 kOe is lifted up above 0.7 GHz (a horizontal solid line). If we change the angle to $\theta = -1°$ (red dotted line) while we fix the frequency 0.7 GHz, a pair of resonances reemerges at high fields.

In order to examine the multiple resonances at 0.7 GHz more closely, we show the microwave absorption spectra for $\theta = -2, -1, 0, 1, 2$ and $4°$ in Fig. 6; here the curves are offset horizontally and vertically for clearer viewing along with a symbol indicating the position predicted by the theoretical calculations. Here we use filled circles, open squares, and open triangles as markers for the primary, secondary and tertiary resonances respectively. We can check the evolution of the spectrum from $\theta = -2°$ to $\theta = 4°$ by following the corresponding f(H) curves in Fig. 5 as previously noted. The clear triple resonance at

$\theta = +2°$ is to be contrasted to the very weak single resonance at $\theta = -2°$, clearly demonstrating the asymmetry. We can also identify three resonances for $\theta = -1, 0, 1°$, as predicted by theory. A double resonance feature was reported earlier for a (001) YIG film by Manuilov et. al.[31], however a triple resonance has never been observed. Unlike the primary resonance peaks, however, secondary and tertiary resonances are broad and hence less distinct. It is interesting to observe the huge change in the FMR spectrum induced by only a one degree change from $\theta = -1°$ to $\theta = -2°$.

**Conclusion**

Frequency-dependent and angle-resolved FMR spectra of a (111) YIG film were measured with a field sweep method using a broadband FMR spectrometer for frequencies in the range 0.7 ~ 8 GHz. By comparing to a theoretical model based on the Liftshitz-Landau formalism that incorporates uniaxial and magnetocrystalline anisotropy through second-order, three anisotropy constants and the saturation magnetization were determined:

$K_1 = -5.421 \times 10^3 \text{ erg/cm}^3 \, (H_{A1} = -82.87 \text{ Oe})$,

$K_2 = 1.372 \times 10^3 \text{ erg/cm}^3 \, (H_{A2} = 20.98 \text{ Oe})$

$K_u = -1.230 \times 10^3 \text{ erg/cm}^3 \, (H_u = -18.80 \text{ Oe})$

$4\pi M = 1.644 \text{ kG}$.

An improvement in accuracy of these constants was accomplished by 1) an enhanced sensitivity of the resonant field to the anisotropy in the low frequency (< 2 GHz) for a near-normal orientation of the field, where the second derivative of the MAE term with respect to the azimuthal angle is comparable to the Zeeman term; and 2) excluding the error associated with determining the azimuthal orientation of the field relative to the in-plane [110] axis of the epitaxial YIG film, which is obtained through the fitting process itself. In addition, multiple resonances observed at low frequencies (0.7 GHz and 1.0 GHz) boost the accuracy of data fitting process. In particular we observed an abrupt "switching-on" of a sharp resonance induced by a less than 1-degree change of the field angle, which can potentially have applications.

## Acknowledgement

This work was supported by the U. S. Department of Energy under grant DE-SC0014424 and the National Science Foundation under grant DMR-1507058. The film growth was supported by the National Natural Science Foundation of China (NSFC) under Grants 51272036 and 51472046. This work was also supported by the National Research Foundation of Korea(NRF) grant funded by the Korean government(MSIP) (NRF-2015R1A4A1041631).

**Figure Captions**

Fig. 1 Spherical coordinate system for the FMR study of an epitaxial YIG film. The film normal is taken as z-axis, which is also the $[111]$ crystal axis; x- and y-axis are parallel to $[2\bar{1}\bar{1}]$ and $[01\bar{1}]$ in crystal axis.

Fig. 2 FMR frequency data as a function of magnetic field normal to the film plane. Data are represented as filled circles while the line shows least squares fit to a straight line.

Fig. 3 Resonant field versus polar field angle at frequencies of 0.7, 1, 1.6, 2.5, 4, 6, and 8 GHz. The experimental data are shown as the various symbols in the legend and the solid lines show a best fit.. Inset shows a magnified view for 0.7, 1.0, and 1.6 GHz near θ =0.

Fig. 4 Fractional changes in resonant field as a function of θ at different frequencies resulting from a 5% change of $k_1$ relative to the best-fit value. Inset shows the same with $\phi_H$

Fig. 5 FMR frequency vs. field for various values of θ. For positive (negative) angles, the data are represented with filled (open) symbols and the theoretical lines are represented by solid (dotted) lines. A horizontal line is drawn to indicate points for 0.7 GHz.

Fig. 6 FMR absorption spectra at 0.7 GHz for $\theta = -2, -1, 0, 1, 2$ and $4°$. For clear viewing, curves are offset both horizontally and vertically. The positions of the primary resonances predicted by the model are denoted by filled circles; open squares and open triangles denote the secondary and tertiary resonances.

**Figures**

**Fig. 1**

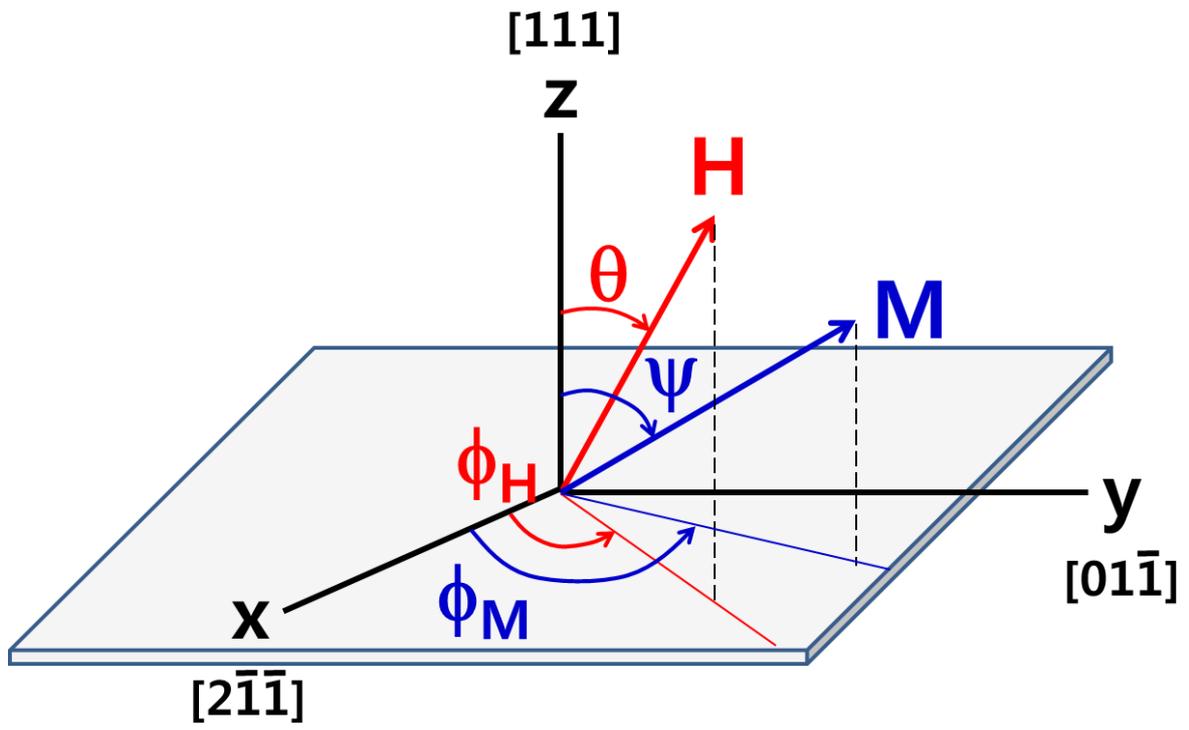

**Fig. 2**

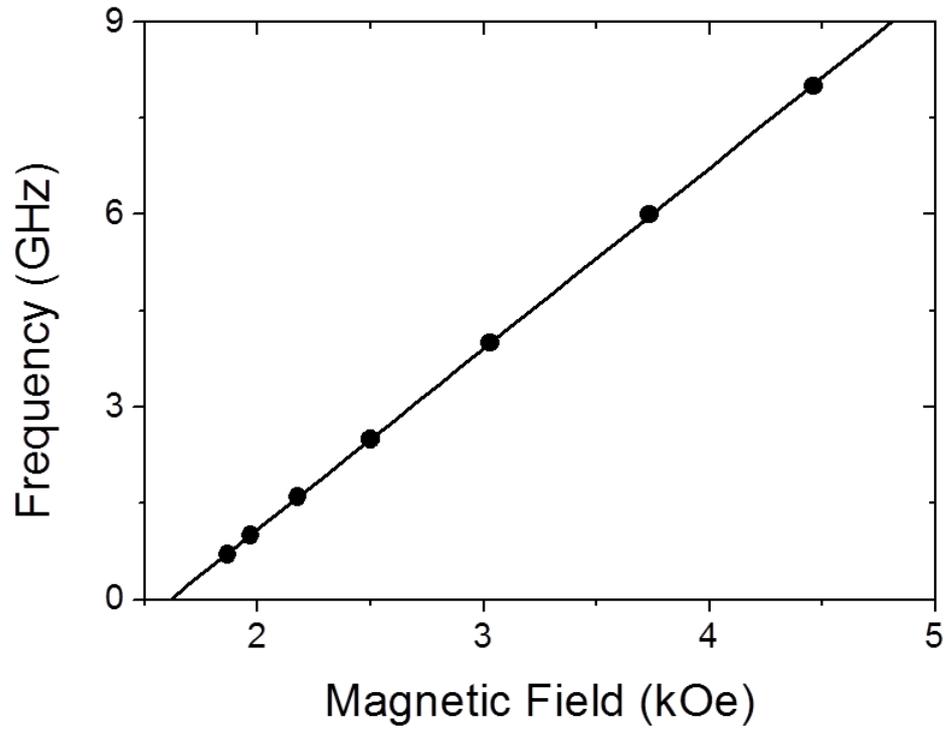

**Fig. 3**

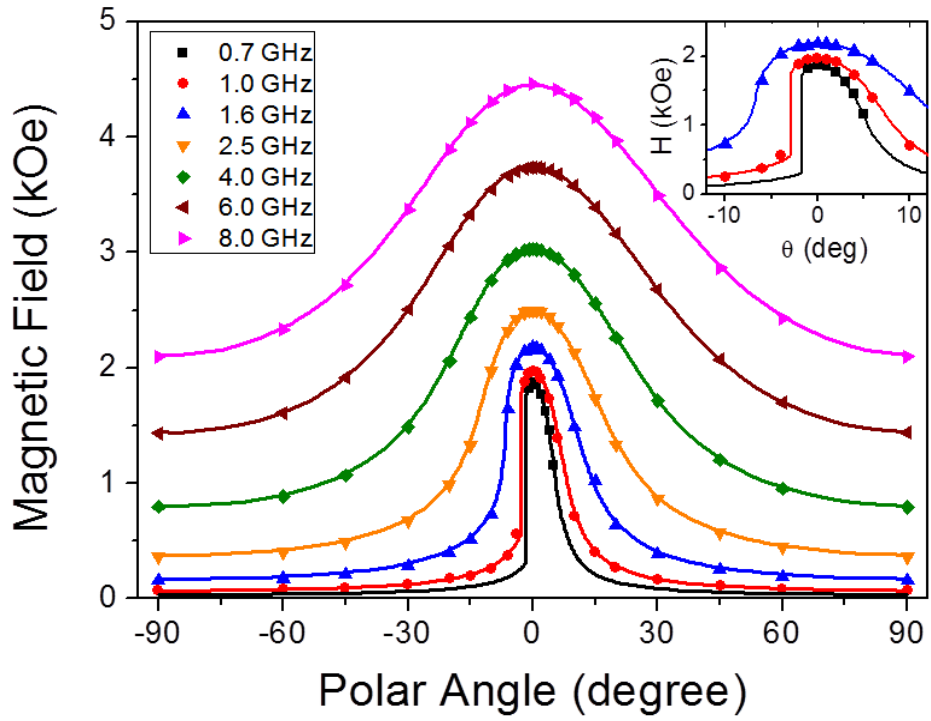

**Fig. 4**

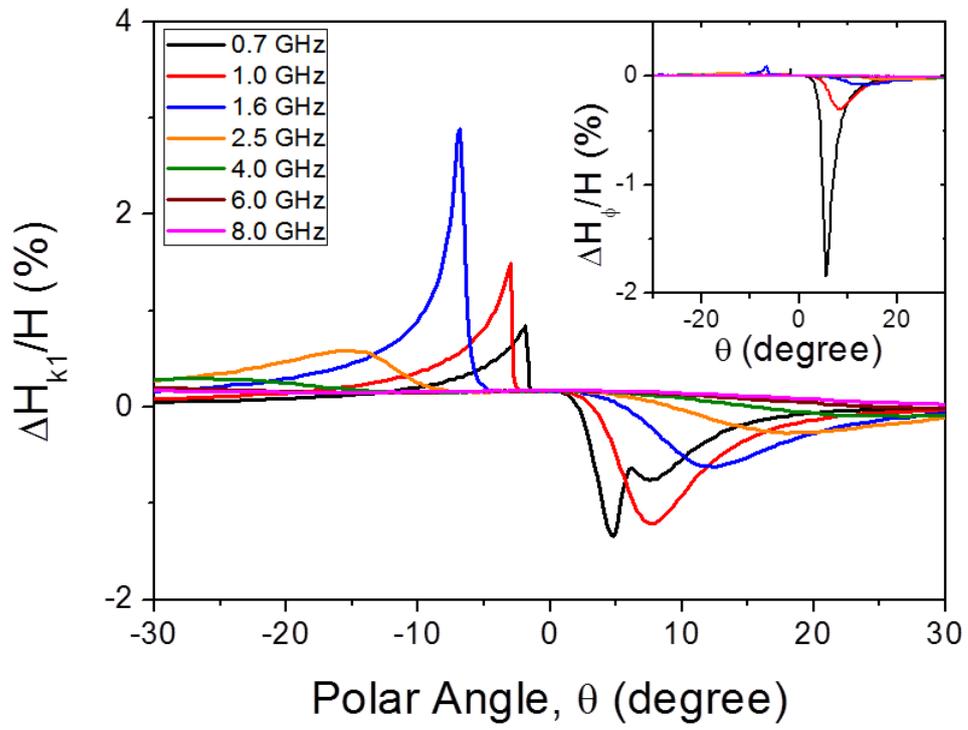

**Fig. 5**

**Fig. 6**